\documentclass[conference]{IEEEtran}
\IEEEoverridecommandlockouts
\usepackage{cite}
\usepackage{amsmath,amssymb,amsfonts}
\usepackage{graphicx}
\usepackage{textcomp}
\usepackage{xcolor}
\usepackage{algorithm}
\usepackage{algorithmic}
\usepackage{url}
\newcommand*{\Comb}[2]{{}^{#1}C_{#2}}  

\def\BibTeX{{\rm B\kern-.05em{\sc i\kern-.025em b}\kern-.08em
    T\kern-.1667em\lower.7ex\hbox{E}\kern-.125emX}}
\begin{document}

\title{Ciphertext-Only Attack on a Secure $k$-NN Computation on Cloud}


\author{\IEEEauthorblockN{Shyam Murthy}
\IEEEauthorblockA{\textit{IISc Bangalore, IN} \\
shyamsm1@iisc.ac.in}
\and
\IEEEauthorblockN{Santosh Kumar Upadhyaya}
\IEEEauthorblockA{\textit{IIIT Bangalore, IN} \\
santosh.upadhyaya@iiitb.ac.in}
\and
\IEEEauthorblockN{Srinivas Vivek}
\IEEEauthorblockA{\textit{IIIT Bangalore, IN} \\
srinivas.vivek@iitb.ac.in}
}

\maketitle

\begin{abstract}
The rise of cloud computing has spurred a trend of transferring data storage and computational tasks to the cloud. To protect confidential information such as customer data and business details, it is essential to encrypt this sensitive data before cloud storage. Implementing encryption can prevent unauthorized access, data breaches, and the resultant financial loss, reputation damage, and legal issues. Moreover, to facilitate the execution of data mining algorithms on the cloud-stored data, the encryption needs to be compatible with domain computation. The $k$-nearest neighbor ($k$-NN) computation for a specific query vector is widely used in fields like location-based services. 
Sanyashi et al. (ICISS 2023) proposed an encryption scheme to facilitate privacy-preserving $k$-NN computation on the cloud by utilizing Asymmetric Scalar-Product-Preserving Encryption (ASPE). 
In this work, we identify a significant vulnerability in the aforementioned encryption scheme of Sanyashi et al. Specifically, we give an efficient algorithm and also empirically demonstrate that their encryption scheme is vulnerable to the ciphertext-only attack (COA).


\end{abstract}


\section{Introduction}
As cloud computing continues to evolve, an increasing number of data owners (DO) are transferring their data to the cloud\cite{b10} \cite{b11}. This shift aids DOs in alleviating the burden associated with data management, computation, and query processing \cite{b1} \cite{b8}. However, the move towards cloud services also raises concerns regarding data security and privacy. Thus, the choice of encryption protocol becomes crucial, especially when computations need to be performed on encrypted data. Traditional encryption techniques, while effective for securing data, do not offer the capability to perform computational operations within the encrypted domain. Traditionally, in order to perform computations on data, the data must first be decrypted, potentially exposing it to security risks. On the other hand, homomorphic encryption schemes \cite{b5} \cite {b6} \cite{b7} provide a potential solution to this issue. These schemes are designed to allow computations to be carried out directly on encrypted data, without the need for decryption. However, despite these potential benefits, the effectiveness of homomorphic encryption schemes for real-world data computations is not entirely certain. These schemes can be complex and computationally intensive, which can limit their efficiency and practicality for large-scale or real-time data computations. Ideally, encryption schemes that  secure the data as well as support search processing would be best suited for this scenario \cite{b13} \cite{b14} \cite{b15} \cite{b16}. Computing $k$ number of nearest neigbours ($k$-NN) of a given query point in a database, according to some metric, is an important technique in the field of machine learning, among others. Privacy-preserving $k$-NN is, hence, equally important in privacy-preserving machine learning (PPML).  Wong et al. \cite{b2} introduced Asymmetric Scalar-product-Preserving Encryption (ASPE),
a scheme that preserves the scalar product ordering between two encrypted data points, when searching an ASPE encrypted database.
While the work of Zhu et al. \cite{b3} encrypted the queries using the Paillier scheme, the work of Sanyashi et al. \cite{b4} 
encrypts the queries in the ASPE scheme itself thereby making the query encryption and the overall scheme more efficient. 
In the scheme of Sanyashi et al. \cite{b4}, encrypting a data tuple involves affine-shifting the individual data items by a secret vector, appending a vector of random nonces, and then multiplying the resulting vector by a secret random matrix results in a ciphertext. The authors argue that such a product vector serving as a ciphertext exhibits a high degree of randomness thereby preserving its security. \\

\noindent {\bf Prior Attacks on the ASPE scheme:} Chunsheng et al. \cite{b17} gave a known-plaintext attack on the original ASPE scheme by solving the ciphertext equations corresponding to known plaintexts.  
Li et al. \cite{b12} used independent component analysis (ICA), which is used for blind source separation in signal processing to show that the original ASPE scheme is not secure against ciphertext-only attacks (COA). 
In this work, we present a COA attack on the scheme of Sanyashi et al. \cite{b4}, where we make use of linear independence properties of differences of encryptions of two data tuples to distinguish between two sets of ASPE-like encrypted ciphertexts (Sec. \ref{sec:COA Attack}). We stress that the attacks
on the original ASPE scheme does \textit{not} necessarily imply attacks on the scheme of Sanyashi et al. \cite{b4} due to ciphertext randomization.
\subsection{Our Contribution}
In this paper, our primary contribution lies in the analysis of the scheme proposed by Sanyashi et al. \cite{b4},
specifically to look at the COA security of the encryption scheme. We prove that the encryption scheme is \textit{not} COA secure. 
It is argued by the authors that the product of a vector, which comprises of affine-shifted data and random nonces, when multiplied with a random secret matrix would yield indistinguishable random ciphertexts. In this work, we revisit the validity of this analysis from the point of view of ciphertexts reflecting differences in the underlying plaintexts, and, hence, find that the assumption is not valid.

We present emperical evidence to suggest that the scheme proposed by Sanyashi et al. \cite{b4} does not provide COA security.  We implement a COA attack on this scheme, demonstrating that the attacker's distinguishing advantage is consistently $\approx 1$, in all trials of a few hundred test runs. This finding substantiates our argument concerning the absence of COA security for the encryption scheme in \cite{b4}. The experimental results are described in Section \ref{sec:result}. Our code is available at  \url {https://github.com/Santosh-Upadhyaya/ICCN-INFOCOM-24/blob/main/coa-attack.ipynb}


\subsection{Organization of the Paper}
In Section \ref{sec:Recap}, we present a recap of the protocol by Sanyashi et al. \cite{b4}. Section \ref{sec:COA Attack} presents the details of the COA indistinguishability game and our COA attack. In Section \ref{sec:result}, we present the details and results of our experiment. Section  \ref{sec:conclusion} concludes the paper. 

\section{Recap of the protocol by Sanyashi et al.} \label{sec:Recap}
The scheme put forth by Sanyashi et al.\cite{b4} serves as an improvement over the one proposed by Zhu et al.\cite{b3}. In the subsequent sections, we will summarize the Key Generation, Encryption, and Decryption components of the scheme from Sanyashi et al.\cite{b4}. The aspects of Query Encryption and Secure $k$-NN computation are not included in this recap as they do not pertain directly to the current study.

\subsection{Key generation} \label{Key generation}
Consider a database $\Delta$ that comprises of $n$ vectors with $d$ dimensions, $n, d \in \mathbb{Z}^+$, the set of positive integers. The elements of the database are assumed to be real numbers.
\\{\em Remark: Throughout this work, we consider real numbers to be sampled uniform randomly and independently from a finite set with suitable bounds, and are represented in a fixed-point scaled integer representation in the underlying hardware architecture.}

The key generation phase has the following steps.  
\begin{itemize} 
\item The public parameters $c$ and $\epsilon$ are generated, where $c, \epsilon \in \mathbb{Z}^+$. 
 
\item A secret vector ${\bf s} \in \mathbb{R}^{d+1}$ is uniform randomly sampled.
\item A secret matrix $\bf{M} \in \mathbb{R}^{\eta \times \eta}$ is uniform randomly sampled.
\item A secret vector $\bf{w} \in \mathbb{R}^c$ is uniform randomly sampled.

\end{itemize}

Finally, ($\bf{s}$, $\bf{M}$, $\bf{w}$) is the secret key for the encryption scheme.

\subsection{Encryption} \label{Data encryption}
During data encryption, DO generates a nonce vector {\bf{z}} of length $\epsilon$. Consider the data to be encrypted be ${\bf{m}}_i$ such that ${\bf{m}}_i$ = $(m_{i,1}, m_{i,2}, \cdots, m_{i,d}) \in \mathbb{R}^d$. Let $\bf{w}$ be a (fixed) secret vector of length $c$, and let $\bf{s}$ be a (fixed) secret vector of length $d+1$ such that the elements of $\bf{s}$ are $s_1$, $s_2$, $\cdots$, $s_{d+1}$. The data is pre-processed (affine shifted by $\bf{s}$) and subjected to encryption by vector multiplication of the pre-processed data with the inverse of a (fixed) secret matrix ${\bf{M}}$ $\in \mathbb{R}^{\eta \times \eta}$, to get a ciphertext ${\bf{c}}_i$ of length $\eta$, where

\begin{multline}\label{encrypt}
{\bf{c}}_i = (s_1 - 2m_{i,1}, \cdots, s_d - 2m_{i, d}, s_{d+1} + \\ ||m_i|| ^ 2, {\bf{w}}, {\bf{z}}_i)) \times \bf{M} ^ {-1}
\end{multline}
The size of each ciphertext vector is $\eta = d + 1 + c + \epsilon$.

\subsection{Decryption}
During the decryption process, 
the nonce vector is first recovered by computing
\begin{equation*}
{\bf{m'}}_i = ({\bf{c}}_i \times \bf{M}),
\end{equation*}
where ${\bf{m'}}_i = (m''{_i}, s_{d+1} + ||m_i|| ^ 2, {\bf{w}}, {\bf{z}}_i)$, and \\
${\bf{m''}}_i = (s_1 - 2m_{i,1}, \cdots, s_d - 2m_{i,d})$.
The individual data elements are recovered as
\begin{equation}
{m_{i,j}}   =  \frac{s_j - m''_{i,j}}{2}.
\end{equation}

\section{COA Attack on the scheme of Sanyashi et al.} \label{sec:COA Attack}
We consider a semi-honest adversarial model in our attack.  The adversary follows the protocol as in the COA indistinguishability game but tries to glean more information than is available per the protocol. 
In this section, we give a quick recap of the COA indistinguishability game, followed by our attack on \cite{b4}.

\subsection{Recap of COA Indistinguishability Game} \label{sec:Recap of COA}
\begin{figure*}[htbp]
  \centering
  \includegraphics[width=\linewidth, height=2.5in]{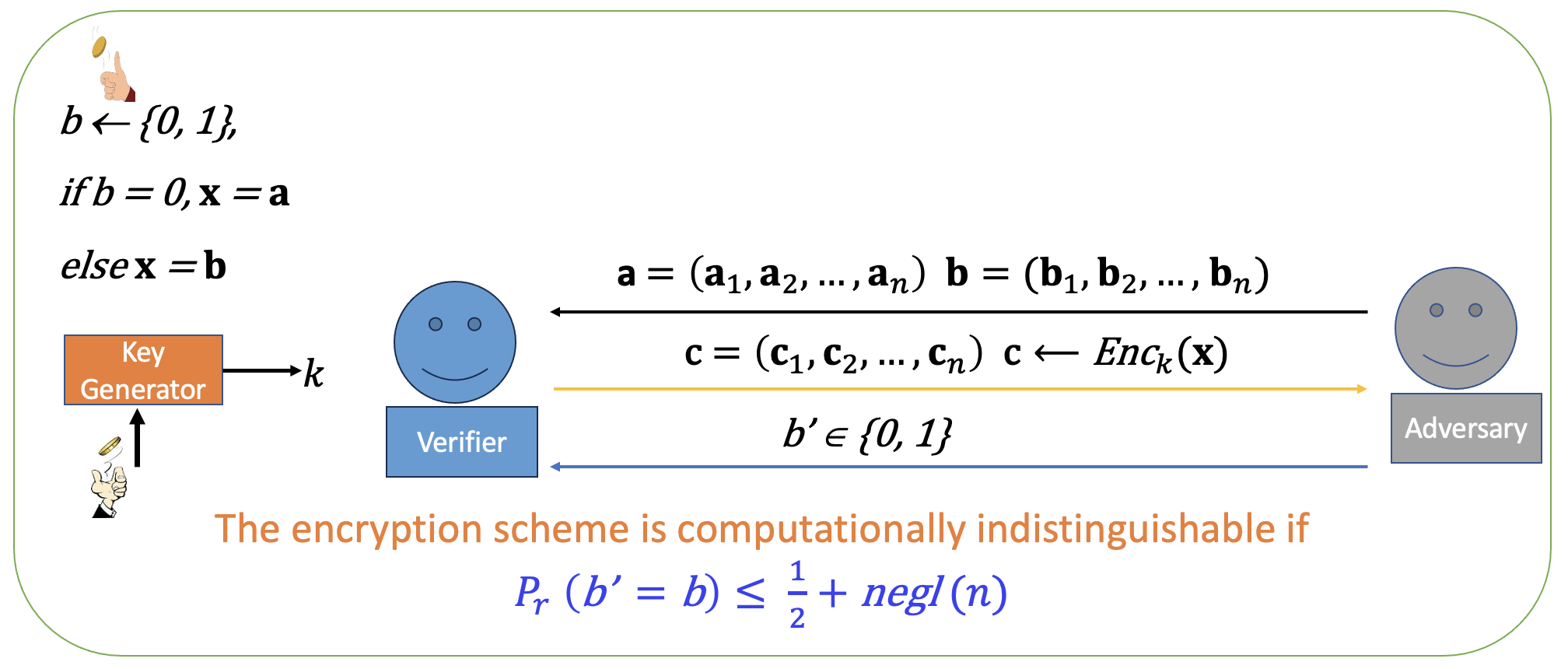}
  \caption{COA Indistinguishibility Game}
  \label{fig:COA-Game-pic}
\end{figure*}

The COA indistinguishability game \cite{b9}, illustrated in Fig. \ref{fig:COA-Game-pic}, involves the following steps:
\begin{itemize}
\item Firstly, the adversary submits two multi-messages (sets of messages), denoted as $\bf{a}$ and $\bf{b}$, to the verifier. 
\item The verifier then randomly selects $b \in \{0,1\}$.
\item Next, the verifier generates the key utilizing the key generation method outlined in Section \ref{Key generation}. The verifier encrypts $\bf{a}$ if $b = 0$, otherwise encrypts $\bf{b}$, using the encryption procedure detailed in Section \ref{Data encryption}.

\item The ciphertexts are then sent to the adversary. 
\item The adversary applies its resources and outputs ${b'}$, which is subsequently returned to the verifier.
\end{itemize}

The encryption scheme is COA secure, if the probability of $(b' = b )$ $\le$ $\frac{1}{2}$ + $negl()$, where $negl()$ is a negligible function  of the security parameter \cite{b18}.

\subsection{Our COA Attack} \label{attack}

The main idea behind our attack is as follows. Since the matrix-vector multiplication is a linear function, the difference of ciphertexts results in an approximate encryption of the difference of the underlying plaintexts with the secrets $\bf{s}$ and $\bf{w}$ now being canceled. Note that if the two message vectors are identical, then the difference of such ciphertexts results in an approximate encryption of the $\bf{0}$ vector with the ciphertext  being the difference of randomly chosen vectors from the subspace spanned by the last $\epsilon$ columns of the secret matrix $\bf{M}$. Note that if the distinguisher sees many (ciphertext) vectors from the subspace, then it can readily recover the basis of the subspace. Using this information, it can then readily determine if a new difference ciphertext belongs to the subspace, and hence, whether it is an approximate ciphertext of the $\bf{0}$ vector. On the contrary, if the underlying message vectors are distinct, then the resulting ciphertext will not be from the above subspace. This is the basis of our COA attack. 

Consider two multi-messages ${\bf{a}} = ({\bf{a}}_1, {\bf{a}}_2, \cdots, {\bf{a}}_n)$ and ${\bf{b}} = ({\bf{b}}_1, {\bf{b}}_2, \cdots, {\bf{b}}_n)$, $n >  \eta$ and each ${\bf{a}}_i$, ${\bf{b}}_i$ $\in \mathbb{R}^d$.  Encryption of $\bf{a}$ (or $\bf{b}$) results in a set $\bf{c_a}$ (or $\bf{c_b}$) of ciphertexts.
The adversary has access to one randomly picked set of ciphertexts $\bf{c} \in \{\bf{c_a}, \bf{c_b}\}$ and let $\bf{c} = (\bf{c_1}, \bf{c_2}, \cdots, \bf{c_n})$. We note here that each individual message encryption, namely ${\bf{c}}_i$, is a vector of size $\eta$. Just as required in the COA indistinguishability game,
the adversary in our attack is given access to one of two sets of ciphertexts, and the semantic security of the underlying cryptosystem is determined by whether the adversary can correctly determine to which of the two multi-messages the set of ciphertexts correspond to.

As per the indistinguishability game, the adversary picks two multi-messages $\bf{a}$ and $\bf{b}$ in the following way and sends them to the verifier. 
Let each individual message, ${\bf{a}}_i  \in \bf{a}$, be a $\bf{0}$ vector of size $d$, and each ${\bf{b}}_i \in \bf{b}$ be a vector of size $d$ with elements drawn randomly from $\mathbb{R}$, such that every ${\bf{b}}_i$ is \textit{distinct} from one another.
As mentioned before, after encryption, the adversary is given only one set of ciphertexts (either the encryption of $\bf{a}$ or that of $\bf{b}$, randomly picked and unknown to the adversary).

{\bf \em Computing Differences of Ciphertexts: } We know that each ${\bf{c}}_i \in  \mathbb{R}^\eta$, is an encryption of a message in $\mathbb{R}^d$.  We pick $\eta$ number of pairs (${\bf{c}}_i$, ${\bf{c}}_j$), ($i > j)$, from the total of $\Comb{n}{2}$ possible pairs with $n > \eta$.  Consider the $\mu \textsuperscript{th}$ pair (${\bf{c}}_i$, ${\bf{c}}_j$), $1\le \mu \le \eta$, in some random ordering, and let \begin{math} {\boldsymbol \delta_\mu} \end{math} = $({\bf{c}}_i - {\bf{c}}_j)$, the difference of ciphertexts.  Similarly, we compute \begin{math} {\boldsymbol \delta_1} \end{math}  to \begin{math} {\boldsymbol \delta_\eta} \end{math}.
W.l.o.g consider one such difference \begin{math} {\boldsymbol \delta_\mu} \end{math}= $({\bf{c}}_i - {\bf{c}}_j)$.  
Let ${\bf{c}}_i, {\bf{c}}_j$ be the encryptions of messages ${\bf{m}}_i, {\bf{m}}_j \in \mathbb{R}^d$, respectively.
In other words, ${\bf{c}}_i$ is the
encryption of ($m_{i,1}, m_{i,2}, \cdots, m_{i,d}$) and 
${\bf{c}}_j$ is the encryption of ($m_{j,1}, m_{j,2}, \cdots, m_{j,d}$).
Then, from Eqn. (\ref{encrypt}), we see that
\begin{multline}
{{\bf{c}}_i} = (s_1 - 2m_{i,1}, \cdots, s_d - 2m_{i, d}, s_{d+1} + \\ ||m_i|| ^ 2, {\bf{w}}, {{\bf{z}}_i}) \times \bf{M} ^ {-1},
\end{multline}
\begin{multline}
{{\bf{c}}_j} = (s_1 - 2m_{j,1}, \cdots, s_d - 2m_{j, d}, s_{d+1} + \\ ||m_j|| ^ 2, {\bf{w}}, {{\bf{z}}_j}) \times \bf{M} ^ {-1}, 
\end{multline}
therefore,
\begin{multline} \label{d_mu}
{\boldsymbol \delta_\mu} = (-2m_{i,1} + 2m_{j,1}, \cdots, -2m_{i, d}+2m_{j,d}, \\ ||m_i||^2 - ||m_j|| ^ 2, {\bf{0}}, {\bf{z}}_i - {\bf{z}}_j) \times \bf{M} ^ {-1}.
\end{multline}
In other words, in each 
\begin{math} {\boldsymbol \delta_i} \end{math}
the secrets {\bf{s}} and {\bf{w}} get canceled,
resulting in the form given 
in Eqn. \ref{d_mu}.  

In the following, we use \begin{math} {\boldsymbol \delta_i} \end{math} to mean a particular instance of the encryption difference, with  $1 \le i \le \eta$.
Recall that \begin{math} {\boldsymbol \delta_i} \end{math} $\in \mathbb{R}^\eta$, $\eta = d+1 + c + \epsilon$, $c=|\bf{w}|$ and $\epsilon=|\bf{z}|$ and {\boldmath $\delta_i$} can be written (from Eqn. \ref{d_mu}) as
\begin{equation} \label{diff_vector}
{\boldsymbol \delta_i} = (r_1, r_2, \cdots, r_d, r_{d+1}, \bf{0}, \bf{z'}) \times \bf{M} ^ {-1} 
\end{equation}
In the case that encryption of $\bf{a}$ was provided to the adversary, then $r_i = 0$, $1 \le i \le d+1$, whereas if the encryption of $\bf{b}$ was provided
to the adversary, then $r_i \in \mathbb{R}, 1 \le i \le d+1$ are expected to be randomly distributed. This is because the corresponding message vectors were randomly chosen, and, hence, their difference is expected to be random.    

{\em {\bf Remark}: Just by looking at the \begin{math} {\boldsymbol \delta_i} \end{math} vectors in Eqn. \ref{diff_vector}, the underlying $r_i$ values remain unknown to the adversary as they are masked by the secret matrix $\bf{M} ^ {-1}$} and $\bf{z'}$.

{\bf \em Description of our attack:} We pick $\epsilon$ number of \begin{math} {\boldsymbol \delta_i} \end{math}s, in no particular order, and form a set {\boldmath $\beta$} = (\begin{math} {\boldsymbol \delta_1} \end{math}, $\cdots$, \begin{math} {\boldsymbol \delta_\epsilon} \end{math}).  We initialize two
counters $in\_span$ and $not\_in\_span$ to 0.  Then, we start with \begin{math} {\boldsymbol \delta_{\epsilon+1}} \end{math}
and check if it is in the span of {\boldmath $\beta$}.  If not, then
we add \begin{math} {\boldsymbol \delta_{\epsilon+1}} \end{math} to {\boldmath $\beta$} and increment $not\_in\_span$ counter.  
Otherwise, we increment $in\_span$ counter.    We repeat this process
for \begin{math} {\boldsymbol \delta_{\epsilon+2}} \end{math} to \begin{math} {\boldsymbol \delta_{3\epsilon}} \end{math}, a total of $2\epsilon-1$ times.  Finally, if $in\_span > \epsilon$, our algorithm
returns $0$, else it returns $1$.

{\bf \em Analysis of our attack:} If the encryption of $\bf{a}$ was provided to the adversary, and since each of the $r_i$ in  Eqn. (\ref{diff_vector}) are 0s, the
first $d+1+c$ elements of \begin{math} {\boldsymbol \delta_i} \end{math} would just be a linear combination of ${\bf z'}$ combined
with the last $\epsilon$ columns of $\bf{M}^{-1}$.  When we consider $\epsilon$ number of linearly independent 
\begin{math} {\boldsymbol \delta_i} \end{math}, they would constitute a basis for the vector space of $\bf{z'}$.  Since we have
$\eta$ number of \begin{math} {\boldsymbol \delta_i} \end{math} and $\eta >> \epsilon$, we will be able to find $\epsilon$ number of linearly independent \begin{math} {\boldsymbol \delta_i} \end{math} vectors with a high probability.   
To simplify this check, we start with the set {\boldmath $\beta$} as described above, and then we verify if every new \begin{math} {\boldsymbol \delta_i} \end{math} vector that we pick is in the span of {\boldmath $\beta$}, which would be the case if {\boldmath $\beta$} had only linearly independent vectors.  Otherwise, we add this vector to {\boldmath $\beta$} and after $\epsilon$ many iterations, we are guaranteed to have at least $\epsilon$ linearly independent vectors in {\boldmath $\beta$} and all subsequent $\epsilon$ many \begin{math} {\boldsymbol \delta_i} \end{math} vectors would be in the span of {\boldmath $\beta$} resulting in $in\_span \ge \epsilon$.

On the other hand, if the adversary were given the encryption of $\bf{b}$, then 
 the first $d$ elements of each \begin{math} {\boldsymbol \delta_i} \end{math} would be random.  Therefore, 
with a high probability each of the \begin{math} {\boldsymbol \delta_i} \end{math} vectors 
are expected to be linearly independent of the vectors of {\boldmath $\beta$}, even if $|${\boldmath $\beta$}$|$= $2\epsilon$, and so they will not be present in the span of {\boldmath $\beta$}.  Hence, with a high probability, we would have $in\_span < \epsilon$ and $not\_in\_span \ge \epsilon$.  Our attack method is given in Algorithm 1. 

Given the fact that a random matrix is used to multiply a vector consisting of
affine-shifted data and nonces, it is natural to expect that the resulting
vectors (ciphertexts in this case) are sufficiently randomized. Similarly, multiplication by a random matrix was expected to provide randomness even when differences in ciphertexts are obtained, thus rendering them indistinguishable from random.
However, our method above shows that this is indeed \textit{not} the case.

\begin{algorithm}
\label{algo1}
\caption{Attack on Encrypted Data}
\begin{algorithmic}[1]
\REQUIRE ${\bf{c}}_1, \cdots, {\bf{c}}_n$ \COMMENT{Ciphertext of a multi-message of length $n$ ($n>\eta>>\epsilon)$}\\
\STATE Initialize $not\_in\_span\_cnt \leftarrow 0$, $in\_span\_cnt \leftarrow 0$
\FOR{$i = 0$ to $n - 1$}
    \STATE ${\bf{\delta}}_i \leftarrow {\bf{c}}_{i+1} - {\bf{c}}_i$ \COMMENT{Compute difference of ciphertexts}
\ENDFOR
\STATE $\beta = \{\}$ 
\FOR{$i = 0$ to $\epsilon - 1$}
        \STATE
        ADD $\delta_i$ to set $\beta$
\COMMENT{Pick $\epsilon$ number of differences}
\ENDFOR\\
\FOR{$i = \epsilon$ to $3\epsilon - 1$}
    \IF {$\delta_i$ in LINEAR\_SPAN($\beta$)}
        \STATE $in\_span\_cnt \leftarrow in\_span\_cnt + 1$
    \ELSE
        \STATE $not\_in\_span\_cnt \leftarrow not\_in\_span\_cnt + 1$\\
        \STATE
        ADD $\delta_i$ to set $\beta$
    \ENDIF
\ENDFOR
\IF{$in\_span\_cnt \ge \epsilon$}
    \RETURN 0
\ELSE
    \RETURN 1
\ENDIF
\end{algorithmic}
\end{algorithm}

\section{Results} \label{sec:result}
The proof-of-concept code for the attack is written in SageMath \cite{b19} to simulate the COA indistinguishability game. Our code consists of Key Generation, Encryption and CoA attack
algorithms as outlined in Sections \ref{sec:Recap} and \ref{attack}.
An auxiliary wrapper function  reads the iteration count which is a user-defined value to mean the number of trials of the experiment to be repeated. In the experiment, values  $d$, $c$, $\epsilon$ are considered as configurable global parameters and $\eta$ = $d + 1 + c + \epsilon$. In each iteration, a choice bit, $\bf{0}$ or $\bf{1}$, is picked uniformly random. If $\bf{0}$ is chosen, a multi-message of size ($\eta + 1$)  consisting of vectors of all $\bf{0}$s is used for encryption, otherwise, a multi-message of size ($\eta + 1$) with distinct vectors with random elements is used for encryption. This part mimics the role of the verifier and outputs the ciphertext.

The attack algorithm obtains the ciphertext from the verifier and executes the method described in Sec. \ref{attack}.
It finally outputs a bit as the result which is returned to the verifier.

{Thorough testing of the code was done with various values of $d$ and we iterated the experiment $512$ times. Our goal was to validate the accuracy of our attack algorithm. This testing involved setting $c = 5$ and $\epsilon = 5$.} The attack algorithm successfully distinguished the ciphertexts with $100\%$ accuracy. As stated in Sec. 2.2 of \cite{b3}, the value of $d$ is expected to be $100$, and our testing was conducted with $d$ values ranging from $8$ to $128$. As given in Sec. 7.1 of \cite{b4}, the security parameters $c$ and $\epsilon$ can be set to $5$,
and our experiments with various $d$ values were conducted with $c = \epsilon = 5$. 
The average execution time for a single attack 
is provided in Table \ref{result-tab}.

\begin{table}[htbp]
\centering
\caption{\textbf{Performance of attack algorithm}}
\label{result-tab}
\begin{tabular} { | c | c |}
\hline
Value of $d$ & Average Execution time of \\
& {attack (sec)} \\
\hline
\hline
$8$ & $0.08$ \\
 \hline
$16$ & $0.09$ \\
 \hline
$32$ & $0.26$ \\
\hline
$64$ & $0.94$ \\
\hline
$128$ & $3.83$ \\
\hline
\end{tabular}
\end{table}

\section{Conclusion} \label{sec:conclusion}
The primary focus of this research was to devise a COA attack on the scheme used in \cite{b4}, based on the multi-message COA indistinguishability game. The proposed attack revealed that the attacker's distinguishing advantage is $\approx 1$. This invalidates the assertions made in Lemma 2 in \cite{b4}. This investigation highlights the necessity for comprehensive security assessments in the design of cryptographic systems. 
It would be interesting to explore modifications to the encryption scheme of \cite{b4} that would thwart our attack. More importantly, such a proposal should be accompanied with a rigorous security analysis based only on the hardness assumptions of well-established problems.

\section*{Acknowledgment}
This work was supported by the Infosys Foundation Career Development Chair Professorship grant for the third author (Srinivas Vivek). 

\footnote{This work has been accepted for publication in the proceedings of the ICCN@INFOCOM 2024 workshop. The final version will be available at https://link.springer.com/.}

\end{document}